\documentclass[preprint,showpacs,preprintnumbers,amsmath,amssymb]{revtex4}

\usepackage{graphicx}
\usepackage{dcolumn}
\usepackage{bm}

\begin{document}

\title{Evidence for divergent baryon-number susceptibility in QCD matter}

\author{N. G. Antoniou}
\email{nantonio@phys.uoa.gr}
\author{F. K. Diakonos}
\email{fdiakono@phys.uoa.gr}
\author{A. S. Kapoyannis}
\email{akapog@phys.uoa.gr}
\affiliation{Department of Physics, University of Athens, 15771 Athens, Greece}

\date{7 July 2008}

\begin{abstract}
The baryon-number density formed in relativistic nuclear collisions, versus the chemical potential of the freeze-out states, is systematically studied on the basis of existing measurements. A remarkable power-law behaviour of the baryon-number susceptibility is found at the SPS, consistent with the existence of a QCD critical point at $\mu_{B,c} \simeq 214$ MeV, $T_c\simeq 155$ MeV. The equation of state in different asymptotic regimes of the critical region is also examined and confronted with freeze-out states in these experiments.
\end{abstract}

\pacs{25.75.Nq,21.65.Qr,12.38.Mh,05.70.Ce}

\maketitle

Baryons produced in collisions of nuclei at relativistic energies play an important role in the identification of critical states in strongly interacting matter \cite{1},\cite{2}. The baryon-number density $n_B(\vec{x})$ is an appropriate order parameter of the critical system, equivalent to the chiral field $\sigma(\vec{x})$ since they both incorporate in their behaviour, the universal power laws and the related exponents of the QCD critical point \cite{1}. A complete set of observables associated either with the baryon-number density or the sigma field, may form together with the corresponding measurements in experiments with nuclei, the basic ingredients and tools for the construction of an observational theory of critical QCD matter \cite{2},\cite{3}. A search for power-law fluctuations in the isoscalar component of $\pi^+\pi^-$ pairs near the two-pion threshold or in the net-baryon system produced at midrapidity, are examples of such observables \cite{2},\cite{3}.

In this Letter we search for a singular behaviour of baryon-number susceptibility $\chi_B$ exploiting the existing measurements of baryochemical potential in a series of freeze-out states ranging from SPS to RHIC energies. The guidelines in this investigation come from the universal properties of the equation of state for 3d Ising systems which specify the universality class of the QCD critical point \cite{4}. The observables in this treatment are (a) the baryon number density $n_B$, properly averaged in a space-time region occupied by the system and (b) the baryochemical potential $\mu_B$ and the temperature $T$ of the corresponding freeze-out state. The behaviour near the critical point is described by the order parameter $\rho_B=n_B-n_{B,c}$ and the conjugate ordering field $m=\frac{\mu_B-\mu_{B,c}}{\mu_{B,c}}$. With these variables the equation of state near the critical point is written as follows \cite{5}:
\begin{equation}
B_c^{\delta}|m|=|\rho_B|^{\delta} f\left(B^{1/\beta}\frac{t}{|\rho_B|^{1/\beta}}\right)\; ; \; t=\frac{T-T_c}{T_c}
\end{equation}

In eq.~(1) the scaling function $f(x)$ is universal and properly normalized: $f(0)=1$ and $f(-1)=0$ \cite{5}. Moreover, the function $f(x)$ is regular at the point $x=0$ and follows a power law, $f(x)\sim x^\gamma$, for large $x$ \cite{5}. The constants $B_c$, $B$ in eq.~(1) are nonuniversal amplitudes and the critical exponents $\beta$, $\gamma$, $\delta$ are fixed by the 3d Ising-QCD universality class ($\beta \simeq \frac {1}{3}$, $\gamma \simeq \frac{4}{3}$, $\delta \simeq 5$).

In the notation of reference \cite{5} the critical equation of state for strongly interacting matter, given by the generic equation (1), is specified as follows \cite{5}:
\begin{subequations}
\begin{equation}
y=f(x)\; ; \; x=\frac{B^{1/\beta} t}{|\rho_B|^{1/\beta}}\; , \; 
y=\frac{B_c^{\delta} |m|}{|\rho_B|^{\delta}}
\end{equation}
\begin{equation}
f(x)=1+\sum_{n=1}^\infty f_n^{(0)} x^n\;\;(x\simeq0)
\end{equation}
\begin{equation}
f(x)=x^\gamma \sum_{n=0}^\infty f_n^{\infty} x^{-2n\beta}\;\;(x\rightarrow\infty)
\end{equation}
\end{subequations}

For 3d Ising systems, the coefficients ($f_n^{(0)}$, $f_n^{\infty}$) have been evaluated in reference \cite{5} for $n\le5$, using renormalization group techniques. The truncation of the series (2b, 2c) at the maximal order $n=5$, leads to a good approximation of $f(x)$ in the limiting regimes ($x\simeq0$, $x\rightarrow\infty$) but also it guarantees a smooth transition from one region to the other around the point $x\simeq1$.

Within the framework of eqs.~(2) one may approach the critical point following distinct paths in the QCD phase diagram: for $t=0$ one moves along the critical isotherm, $|\rho_B|=B_c |m|^{1/\delta}$, and the baryon-number susceptibility,
$\chi_B=\frac{\partial n_B}{\partial \mu_B}$, develops a power-law singularity for $m=0$, $\chi_B \sim |m|^{-\varepsilon}$, a clear signal of a critical point ($\varepsilon=\frac{\delta-1}{\delta}$). For $t<0$, $m=0$ we move along the coexistence line, $|\rho_B|=B |t|^{\beta}$, which is a limiting curve of the family (2a) when $m \rightarrow 0$. In the general case ($m \neq 0$) eq.~(2a) penetrates the crossover regime for $t>0$ with a smooth dependence of $\rho_B$ on $t$.
Finally, in the limit $x\rightarrow\infty$ the system follows the critical isochor ($\rho_B\rightarrow0$) along which the baryon-number susceptibility obeys the power-law $\chi_B=C_\chi |t|^{-\gamma}$ with the amplitude $C_\chi
= \frac{B_c^\delta}{f_0^{\infty}B^{\gamma/\beta}}$.
This landscape of universal power laws, extracted from exact equation of state, may be confronted with measurements of the observables: $n_B$, $\mu_B$, $T$ in relativistic nuclear collisions, in order to capture the critical region of strongly interacting matter.

For the net-baryon density $n_B$ we have used two different averaging scales in rapidity in order to verify that the nature of critical phenomena incorporated in the equation of state (2) does not depend on this scale. The first choice corresponds to the total rapidity length $\Delta$ available in the collision and leads to the representation: $n_B(\Delta)=A_w A_{min}^{-2/3} (2\sinh \frac{\Delta}{2})^{-1}$ where $A_w$ is the number of wounded nucleons (participants). The other, rather extreme, choice is a scale of homogeneity at midrapidity ($\delta y \simeq 2$) leading to the expression: 
$n_B(0)= A_{min}^{-2/3}\left(\frac{dN_B}{dy}\right)_0$ where $\left(\frac{dN_B}{dy}\right)_0$ denotes the baryon number per unit of rapidity, in the central region.

In Table \ref{tab:table1} we summarize the experiments with nuclei (A+$\rm A'$) and the corresponding measurements of the relevant observables, in a wide range of energy and size of the colliding systems. The baryonic densities have been calculated using either the formula \cite{PbPb158-data}
\begin{equation}
B-\bar{B}=
(2.07\pm0.05)(p-\bar{p})+(1.6\pm0.1)(\Lambda-\bar{\Lambda})
\end{equation}
or the formula \cite{QM99-Cooper}
\begin{eqnarray}
B-\bar{B}=(1+a)(p-\bar{p})+2\frac{1+b}{1+2b}(K^+-K^-),\nonumber \\
a=1.07,\;b=0.1
\end{eqnarray}

\begin{table*}
\caption{\label{tab:table1} Measurements of the relevant observables in experiments with nuclei.}
\begin{ruledtabular}
\begin{tabular}{ccccccc}
A+$\rm A'$ & $\sqrt{s}(GeV)$ & $A_w$ & $\left(\frac{dN_B}{dy}\right)_0$ & $\mu_B$(MeV) & $T(MeV)$   & Ref. \\ \hline
(S+S)$_{cen.}$             & 19.4                  & 54$\pm$3             &    8.6$\pm$1\footnotemark[1]  & 220$\pm$22            & 180.5$\pm$10.9       &
\cite{NA35-data1},\cite{NA35-data2},\cite{SPS-tha1}  \\ 
(S+Ag)$_{cen.}$            & 16.3                  & 90$\pm$10            &
15$\pm$1\footnotemark[1]   & 242$\pm$18            & 178.9$\pm$8.1        & 
\cite{NA35-data2},\cite{SPS-tha1}  \\
(S+Au)$_{cen.}$            & 13.5                  & 113                  &
22$\pm$2\footnotemark[1]   & 175$\pm$5             & 165$\pm$5            &
\cite{NA35-data1},\cite{NA35-data2},\cite{SPS-tha2}  \\
(Pb+Pb)$_{cen.}$           & 8.77                  & 349$\pm$5            & 108.0$\pm$2.6\footnotemark[2], 106.2$\pm$3.2\footnotemark[1] & 381.6$\pm$6.7         & 144.6$\pm$2.3        &
\cite{PbPb4080-1},\cite{PbPb4080-2},\cite{PbPb-tha1}   \\
(Pb+Pb)$_{cen.}$           & 12.3                  & 349$\pm$5            & 84.8$\pm$2.3\footnotemark[2], 77.2$\pm$2.8\footnotemark[1] & 296.0$\pm$6.4         & 151.7$\pm$2.9        &
\cite{PbPb4080-1},\cite{PbPb4080-2},\cite{PbPb-tha1}   \\
(Pb+Pb)$_{cen.}$           & 17.3                  & 352$\pm$12           & 67.7$\pm$7.6\footnotemark[1]               & 247.4$\pm$5.7         & 156.1$\pm$1.6        &
\cite{PbPb158-data},\cite{PbPb-tha1}  \\
(Pb+Pb)$_{0-5\%}$      & 17.3                  & 362$\pm$12           & 
82.7\footnotemark[2]       & 248.9$\pm$8.2         & 157.5$\pm$2.2        &
\cite{QM99-Cooper},\cite{QM99-Sikler},\cite{PbPb-tha2}  \\
(Pb+Pb)$_{5-14\%}$     & 17.3                  & 304$\pm$16           & 
67.4\footnotemark[2]       & 235.2$\pm$8.5         & 150.6$\pm$3.2        &
\cite{QM99-Cooper},\cite{QM99-Sikler},\cite{PbPb-tha2}  \\
(Pb+Pb)$_{14-23\%}$    & 17.3                  & 241$\pm$16           & 
50.8\footnotemark[2]       & 223.7$\pm$9.9         & 156.6$\pm$3.7        &
\cite{QM99-Cooper},\cite{QM99-Sikler},\cite{PbPb-tha2}  \\
(Pb+Pb)$_{23-31\%}$    & 17.3                  & 188$\pm$16           & 
35.7\footnotemark[2]       & 210$\pm$11            & 154.7$\pm$4.2        &
\cite{QM99-Cooper},\cite{QM99-Sikler},\cite{PbPb-tha2}  \\
(Pb+Pb)$_{31-48\%}$    & 17.3                  & 130$\pm$14           & 
21.6\footnotemark[2]       & 213$\pm$16            & 153.2$\pm$5.9        &
\cite{QM99-Cooper},\cite{QM99-Sikler},\cite{PbPb-tha2}  \\
(Si+Si)$_{cen.}$           & 17.3                  & 56.1\footnotemark[3] & 
-                          & 253$\pm$11            & 162.7$\pm$4.1        &
\cite{PbPb-tha1}           \\
(C+C)$_{cen.}$             & 17.3                  & 24\footnotemark[3]   & 
-                          & 256$\pm$13            & 166.1$\pm$4.3        &
\cite{PbPb-tha1}           \\
(Au+Au)$_{cen.-per.}$      & 130                   & 381-4.9              & 
(0.090-0.018)\footnotemark[1]                & 28.6-17.6             & 165                  &
\cite{STAR-1},\cite{STAR-2},\cite{STAR-3},\cite{PHOBOS}        \\
\end{tabular}
\end{ruledtabular}
\footnotetext[1]{Calculated using $\Lambda-\bar{\Lambda}$.}
\footnotetext[2]{Calculated using $K^+-K^-$.}
\footnotetext[3]{It corresponds to the assumption $A_w=2A$.}
\end{table*}

\begin{figure}
\includegraphics[scale=0.91]{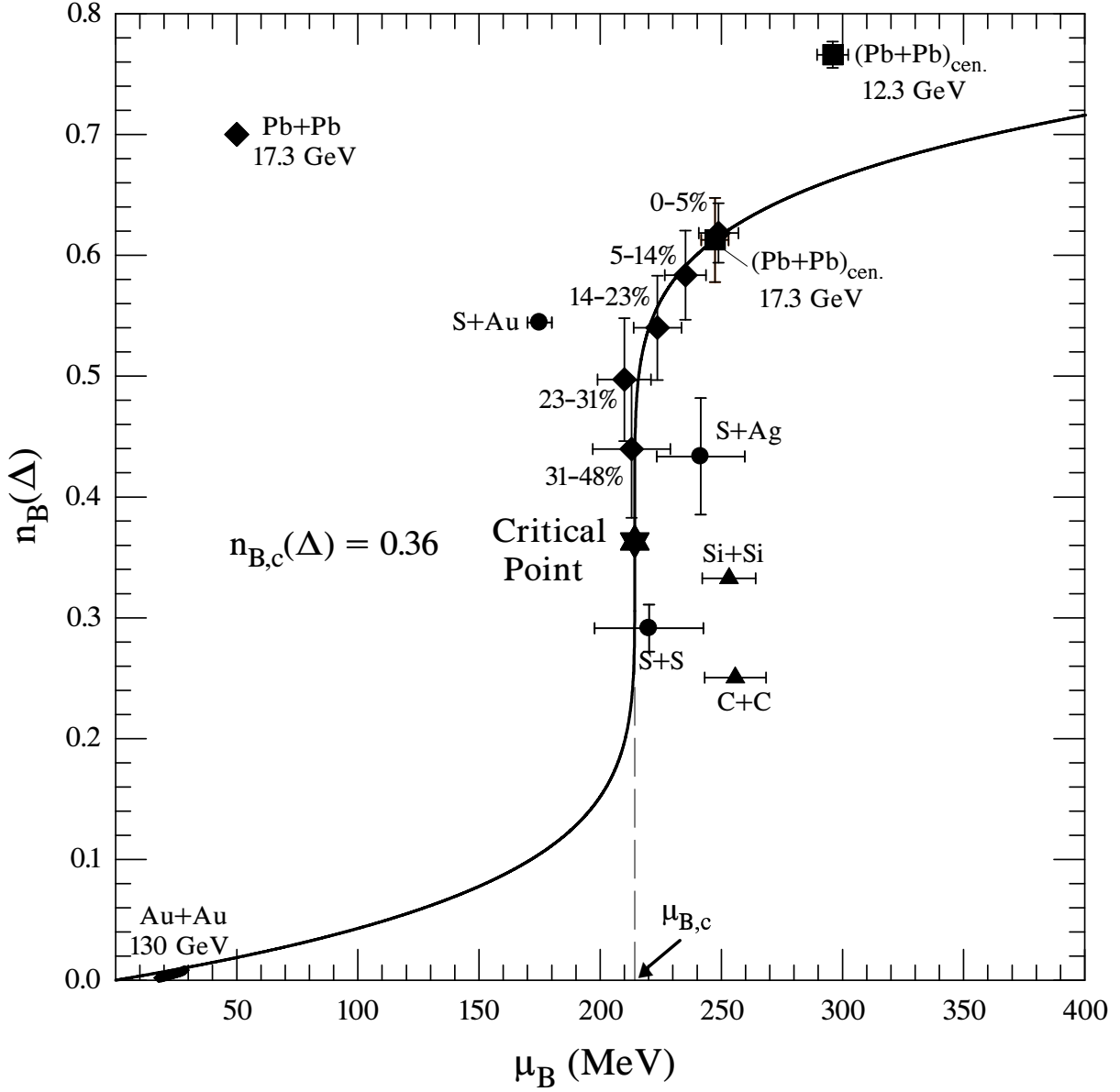}
\caption{\label{fig:rhommb} The baryon number density $n_B(\Delta)$ versus baryochemical potential $\mu_B$ is illustrated for a series of freeze-out states. The solid line represents the best fit solution in the description of the 
Pb+Pb ($\sqrt{s}=$17.3 GeV) central and peripheral data with the critical isotherm equation of state.}
\end{figure}

\begin{figure}
\includegraphics[scale=0.91]{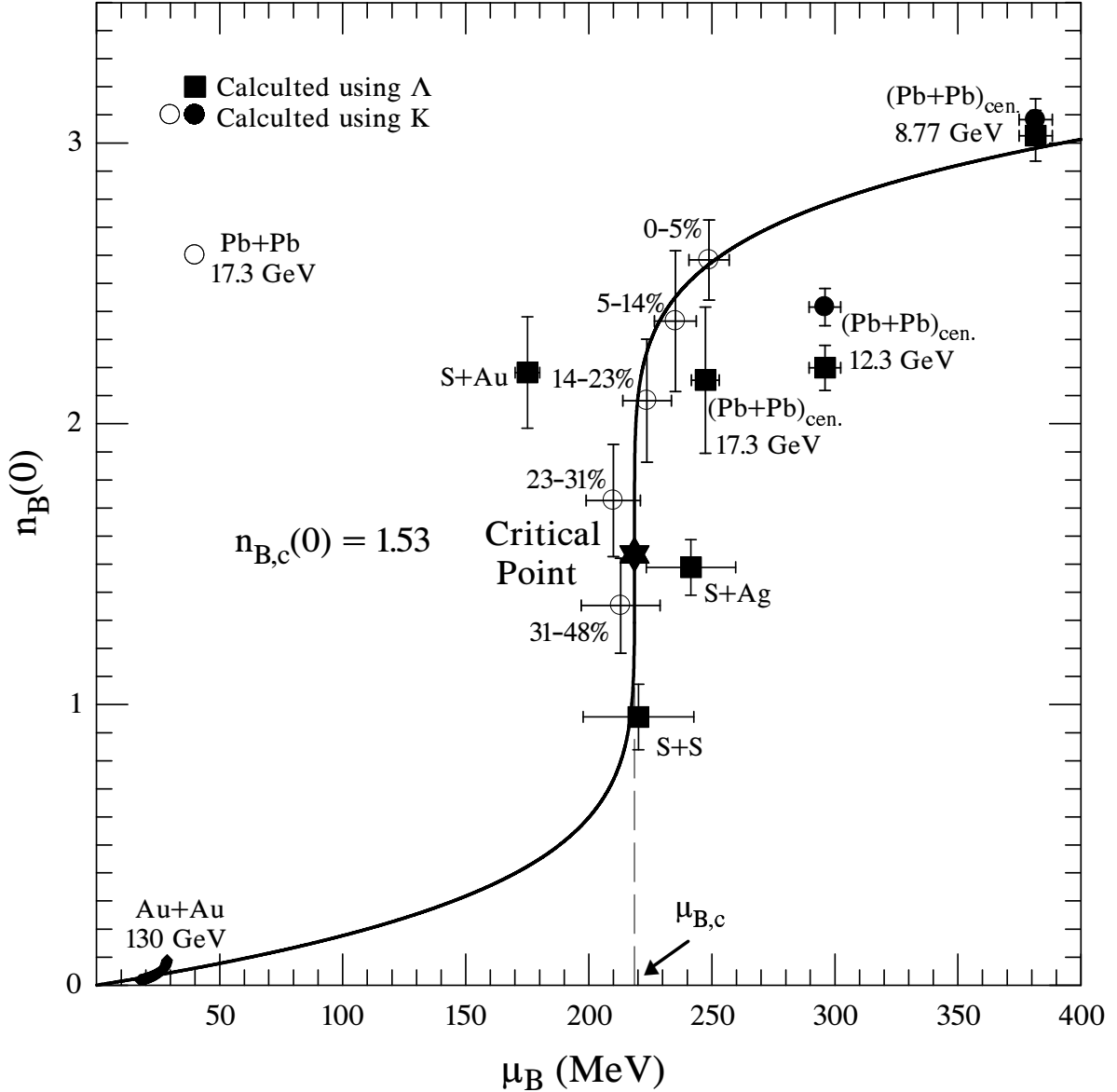}
\caption{\label{fig:rho0mb} The same illustration as in Fig.~\ref{fig:rhommb}, using for $n_B(0)$ the averaging scale at midrapidity. The solid line is produced through a best fit on the Pb+Pb ($\sqrt{s}=$17.3 GeV) central and peripheral data (open circles).}
\end{figure}

\begin{figure}
\includegraphics[scale=0.91]{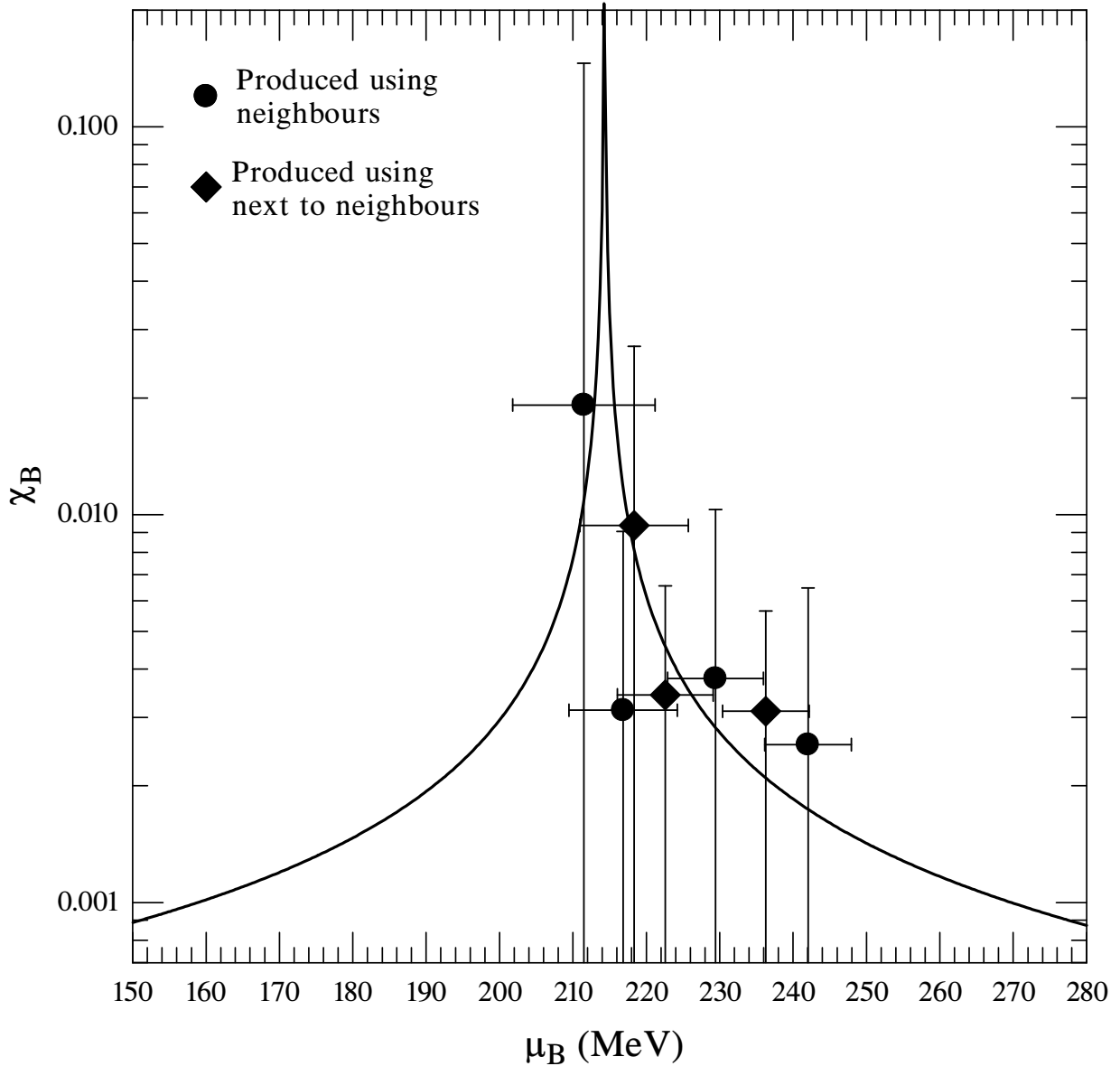}
\caption{\label{fig:drdmmb} The baryon-number susceptibility $\chi_B$, for $T=T_c$ versus $\mu_B$. The solid line corresponds to the derivative of the best fit solution shown in Fig.~\ref{fig:rhommb}. The data points are produced using neighbouring  (cycles) or next to neighbouring (squares) differences.}
\end{figure}

\begin{figure}
\includegraphics[scale=0.91]{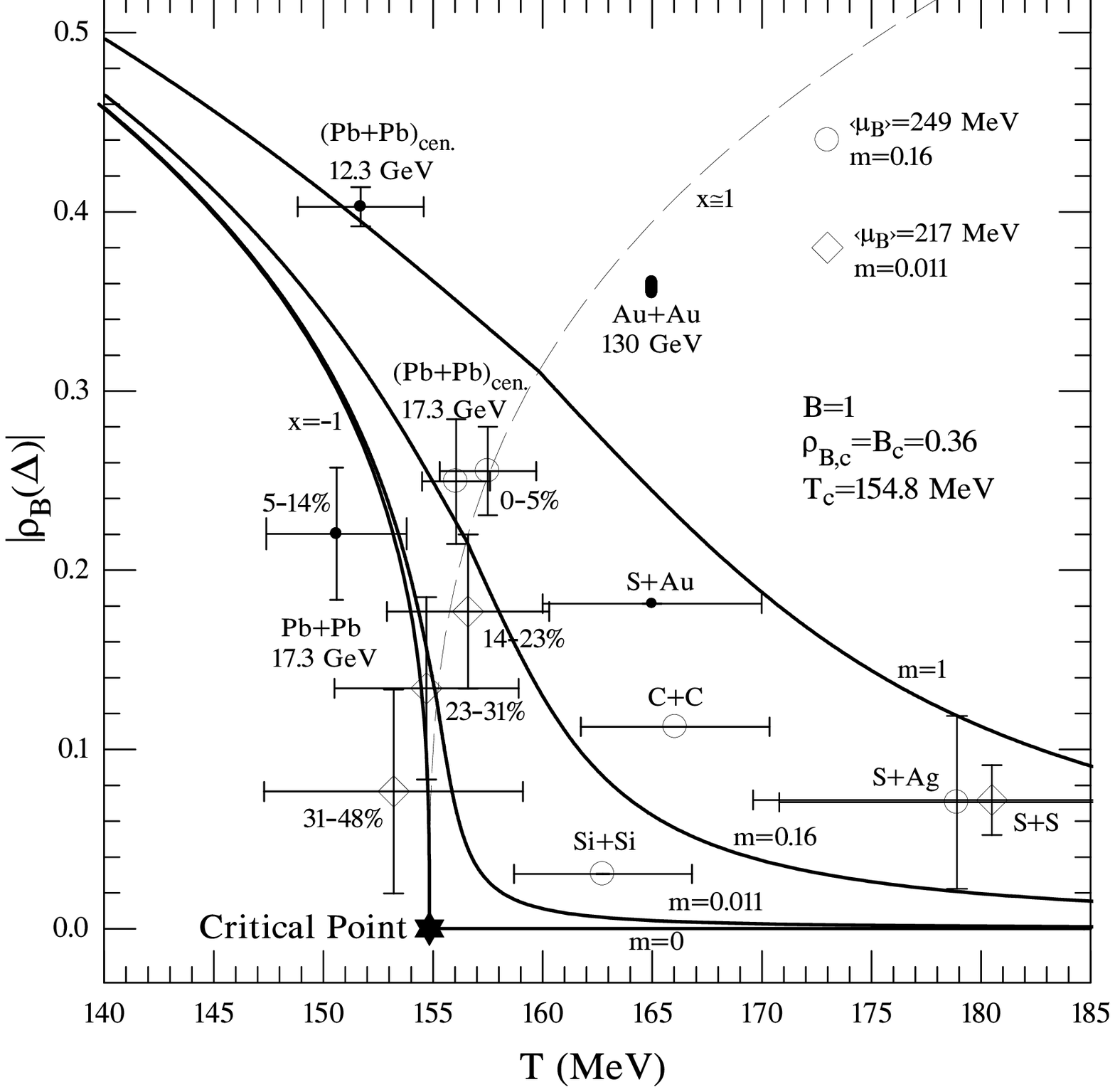}
\caption{\label{fig:rho0T} The equation of state $\left|\rho_B\right|$ versus $T$, for $m=const$. The curve $m=0$, $T\le T_c$ corresponds to the coexistence line and for $T\ge T_c$ to the critical isochor. The experimental freeze-out states are located within the area spanned by the family of curves of constant $m$.}
\end{figure}

In Fig.~\ref{fig:rhommb} the freeze-out states in experiments at the SPS and RHIC are located in the diagram $n_B(\Delta)$ versus $\mu_B$. The experimental points belong to three distinct classes of measurements: (a) central and peripheral Pb+Pb collisions at the SPS (NA49 experiments), (b) central collisions of light nuclei (S, Si, C) at the SPS (NA35 and NA49 experiments) and (c) central and peripheral Au+Au collisions at RHIC. In the states of class (a) and in particular for the highest SPS energy ($\sqrt{s} \simeq 17$ GeV), the freeze-out temperature remains practically constant (Table \ref{tab:table1}) and the actual value measured in these experiments is fixed, on the average, at the level $\left<T\right> \simeq 155$ MeV. As a result we may assume that the experimental points of class (a) belong to an isotherm which, as shown in Fig.~1, develops a very large derivative $\frac{\partial  n_B}{\partial \mu_B}$ near the value $\mu_B \simeq 200$ MeV of the chemical potential. This observation implies that the baryon-number susceptibility of strongly interacting matter may become infinitely large in this region of the phase diagram ($T \simeq 155$ MeV, $\mu_B \simeq 200$ MeV). It is therefore suggestive that the freeze-out states in class (a) form a critical isotherm associated with the 3d Ising-QCD critical point.

In order to verify quantitatively this hypothesis, starting from the above observations, one must employ the critical equation of state discussed in the first part of this Letter (eqs.~(1) and (2)), in the case $t=0$ (critical isotherm). In other words, one must be able to describe the freeze-out states of class (a) in the diagram of Fig.~\ref{fig:rhommb} with the equation of state:
\begin{equation}
\left|n_B(\Delta) - n_{B,c} \right|=
B_c \left|\frac{\mu_B-\mu_{B,c}}{\mu_{B,c}}\right|^{\frac{1}{\delta}}
\end{equation}
With the normalization condition $n_B=0$ for $\mu_B=0$, the best fit solution ($\chi^2/dof \simeq0.16$) is illustrated in Fig.~1 and corresponds to the actual values of the parameters: $n_{B,c}=0.36$, $\mu_{B,c} \simeq 214$ MeV ($B_c\simeq n_{B,c} $).

In Fig.~\ref{fig:rho0mb} the same procedure is applied for a different averaging scale in rapidity, corresponding to a baryon-number density $n_B(0)$, as explained in the first part of this Letter. The critical solution is not affected by this change ($\mu_{B,c} \simeq218$ MeV, $\chi^2/dof \simeq 0.51$) and the overall outcome of this treatment seems to be robust against the change of the averaging scale in rapidity. In what follows we use, therefore, $n_B(\Delta)$ as a typical baryon-number density, on the average.

In summary, we have been able to isolate a class of freeze-out states corresponding to central and peripheral Pb+Pb collisions at $\sqrt{s}\simeq 17$ GeV which form a critical isotherm associated with a QCD critical point at a location in the phase diagram fixed by the values: $T_c\simeq155$ MeV, $\mu_{B,c} \simeq 214$ MeV. The baryon-number susceptibility, $\chi_B \sim |\mu_B-\mu_{B,c}|^{\frac{1-\delta}{\delta}}$, diverges at the critical point according to a universal power law, fixed by the critical exponent $\delta$ (eq.~(5)). In Fig.~\ref{fig:drdmmb}, the singularity of $\chi_B$ corresponding to solution (5) is illustrated together with the results of a crude differentiation of nearby experimental points in Fig.~1. The errors are large but the trend is consistent with the derivative of the solution (5).

In this last part of our treatment we attempt to exploit, beyond the critical isotherm, the universal properties of the critical equation of state which are rigorously incorporated in the representation (2). To this end, in Fig.~\ref{fig:rho0T} we consider the diagram $|\rho_B|$ versus $T$ in which the freeze-out states are properly located together with the critical point found in our approach. With the help of eqs.~(2) we have constructed, in the same figure, the family of curves corresponding to different values of $m$ and representing the critical equation of state in its full capacity (critical surface). Across this family we have drawn the curve $\chi \simeq 1$ (eq.~2a) which, as it was already explained, corresponds to a smooth transition from the representation (2b) to the expansion (2c) of the equation of state. 
For $m=0$, $T\le T_c$, the coexistence curve has a normal scaling behaviour near the critical point, $|\rho_B|\sim|t|^\beta$, and for $B=1$ a good description of the peripheral freeze-out states in class (a) is obtained along this curve. In general the lines of constant $m$, in Fig.~\ref{fig:rho0T}, provide us with a guidance for the identification of the experimental freeze-out states as critical states. In fact the cluster of experimental points in the area $|m|\le0.16$, corresponding to a domain in chemical potential $214\le\mu_B\le250$ MeV, is an example of freeze-out systems consistent with the critical equation of state. 
In particular, the collision Si+Si ($\sqrt{s}\simeq 17$ GeV) freezes out close to the critical point (Fig.~\ref{fig:rho0T}) and as a consequence, power-law density fluctuations are expected to occur in this process \cite{bb.last}. On the other hand, the freeze-out system at RHIC (Au+Au, $\sqrt{s}\simeq 130$ GeV) violates strongly the critical equation of state and therefore a drastic change of the collision parameters is needed (lower energy, smaller systems) in order to capture freeze-out states consistent with criticality \cite{b.last}.

In conclusion, we have shown that the universal form of the critical equation of state for systems belonging to the 3d Ising universality class \cite{5} can be fully exploited in experiments with nuclei in order to identify critical QCD states at the freeze-out regime. In this approach we have found that existing measurements of $n_B$, $T$, $\mu_B$ at SPS energies, reveal, within experimental errors, a class of freeze-out states belonging to the critical isotherm of the QCD universality class and leading, as a consequence, to a divergent baryon-number susceptibility. This phenomenon is a strong indication for the existence of a QCD critical point in the neighbourhood of these states ($\mu_{B,c} \simeq 214$ MeV, $T_c \simeq 155$ MeV). The overall analysis suggests that new, high precision measurements in experiments with light nuclei (S+S, Si+Si, C+C) are needed (SPS-NA61, RHIC at low energies) in order to discover the exact location of the QCD critical point in the phase diagram \cite{b.last},\cite{last}.

\end{document}